\documentclass[twoside,leqno,twocolumn]{article}
\pdfoutput=1
\usepackage{ltexpprt}
\usepackage{pslatex}
\usepackage[T1]{fontenc}
\usepackage[latin1]{inputenc}
\usepackage{amssymb}
\usepackage{graphicx}
\makeatletter
 \usepackage{verbatim}
\usepackage{ifpdf}
\ifpdf
\usepackage[pdftex,plainpages=false]{hyperref}
\fi
\begin{document}

\title{Slope One Predictors for Online Rating-Based Collaborative Filtering}
%\conferenceinfo{Submitted to CIKM}{'04 Washington, D.C. USA}
%\CopyrightYear{2004}
\author{Daniel Lemire\thanks{Université du Québec à Montréal} \\
\and
Anna Maclachlan\thanks{Idilia Inc.}}
\date{}
%\author{Daniel Lemire\inst{1} \and Anna Maclachlan\inst{2}}
%\institute{Université du Québec \and Idilia Inc.}
% \numberofauthors{2}
%  \author{
%  \alignauthor Daniel Lemire\\
%         \affaddr{Université du Québec}\\
%     \affaddr{Montréal, QC, Canada}\\
%         \email{daniel\_lemire@teluq.uquebec.ca}
%  \alignauthor Anna Maclachlan\\
%         \affaddr{National Research Council of Canada}\\
% %%        \affaddr{Institute for Information Technology}\\
%  %%       \affaddr{46 Dineen Drive}\\
%         \affaddr{Fredericton, NB, Canada}\\
%         \email{anna.maclachlan@nrc.gc.ca}
%  }

\maketitle
\begin{abstract}
%\small\baselineskip=9pt
Rating-based collaborative filtering is the process of predicting
how a user would rate a given item from other user ratings. We
propose three related slope one schemes with predictors of the
form $f(x) = x + b $, which precompute the average difference
between the ratings of one item and another for users who rated
both. Slope one algorithms are easy to implement, efficient to
query, reasonably accurate, and they support both online queries
and dynamic updates, which makes them good candidates for
real-world systems. The basic \textsc{slope one} scheme is
suggested as a new reference scheme for collaborative filtering.
By factoring in items that a user liked separately from items that
a user disliked, we achieve results competitive with slower
memory-based schemes over the standard benchmark EachMovie and
Movielens data sets while better fulfilling the desiderata of CF
applications.
\end{abstract}

% \category{H.3.3}{Information Storage and Retrieval}{Information
% Search and retrieval}[Information Filtering]

% \begin{keywords}
% G.4.a Algorithm design and analysis,
% J.8.e Electronic commerce
% \end{keywords}


\renewcommand{\thefootnote}{}
\footnotetext{In SIAM Data Mining (SDM'05), Newport Beach, California, April 21-23, 2005.}
\renewcommand{\thefootnote}{\arabic{footnote}}

\section{Introduction}
An online rating-based Collaborative Filtering CF query consists of an array of
(item, rating) pairs from a single user. The response to that query
is an array of predicted (item, rating) pairs for those items the
user has not yet rated. We aim to provide robust CF schemes that are:

\begin{enumerate}
\item easy to implement and maintain: all aggregated data should be easily interpreted by the
average engineer and algorithms should be easy to implement and test;
\item updateable on the fly: the addition of a new rating should change all predictions instantaneously;
\item efficient at query time: queries should be fast, possibly at the expense of storage;
\item expect little from first visitors: a user with few ratings should receive valid recommendations;
\item accurate within reason: the schemes should be competitive with the most accurate schemes,
but a minor gain in accuracy is not always worth a major sacrifice
in simplicity or scalability.
\end{enumerate}

Our goal in this paper is not to compare the accuracy of a wide
range of CF algorithms but rather to demonstrate that the Slope One
schemes simultaneously fulfill all five goals.
In spite of the fact that our schemes are simple, updateable, computationally
efficient, and scalable, they are comparable in accuracy to schemes
that forego some of the other advantages.

\begin{figure}
% this assumes pdflatex
\begin{center} \includegraphics[height=3.5cm, angle=0]{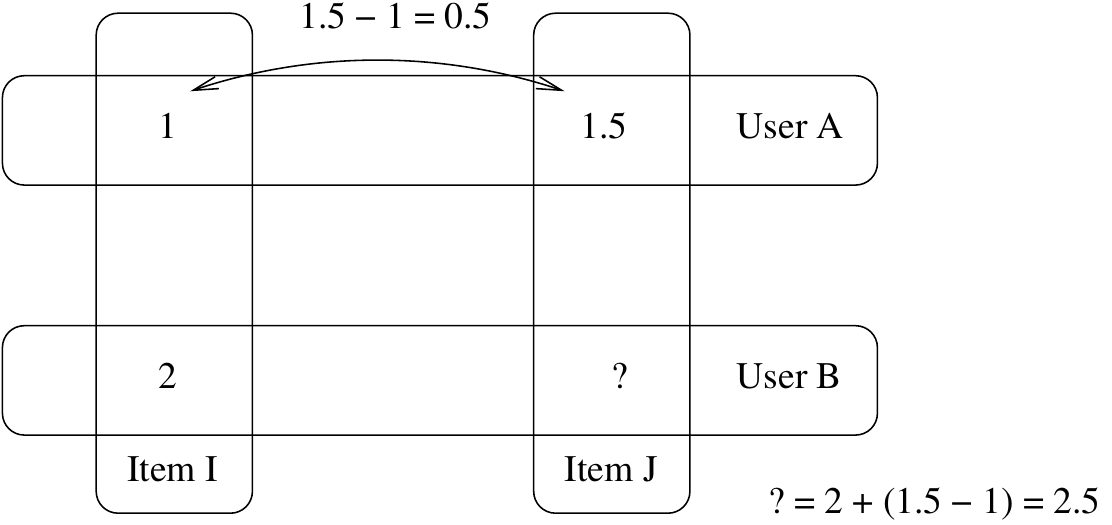}
\end{center}
\caption{\label{figsimple} Basis of \textsc{Slope One} schemes:
User A's ratings of two items and User B's rating of a common item
is used to predict User B's unknown rating.}
\end{figure}

Our Slope One algorithms work on the intuitive principle of a
``popularity differential'' between items for users. In a pairwise
fashion, we determine how much better one item is liked than
another. One way to measure this differential is simply to
subtract the average rating of the two items. In turn, this
difference can be used to predict another user's rating of one of
those items, given their rating of the other. Consider two users
$A$ and $B$, two items $I$ and $J$ and Fig.~\ref{figsimple}. User
$A$ gave item $I$ a rating of $1$, whereas user $B$ gave it a
rating of $2$, while user $A$ gave item $J$ a rating of $1.5$. We
observe that item $J$ is rated more than item $I$ by $1.5 - 1 =
0.5$ points, thus we could predict that user $B$ will give item
$J$ a rating of $2 + 0.5 = 2.5$. We call user $B$ the predictee
user and item $J$ the predictee item. Many such differentials
exist in a training set for each unknown rating and we take an
average of these differentials.
 The family of slope one
schemes presented here arise from the three ways we select the
relevant differentials to arrive at a single prediction.

The main contribution of this paper is to present slope one CF
predictors and demonstrate that they are competitive with
memory-based schemes having almost identical accuracy, while being
more amenable to the CF task.

\section{Related Work}
\subsection{Memory-Based and Model-Based Schemes}
Memory-based collaborative filtering uses a similarity measure
between pairs of users to build a prediction, typically through a
weighted
average~\cite{Empirical,Personality,GroupLens1994,Lightweight}.
The chosen similarity measure determines the accuracy of the
prediction and numerous alternatives have been
studied~\cite{NonPersonalized}. Some potential drawbacks of
memory-based CF include scalability and sensitivity to data
sparseness.  In general, schemes that rely on similarities across
users cannot be precomputed for fast online queries. Another
critical issue is that memory-based schemes must compute a
similarity measure between users and often this requires that some
minimum number of users (say, at least 100 users) have entered
some minimum number of ratings (say, at least 20 ratings)
including the current user. We will contrast our scheme with a
well-known memory-based scheme, the Pearson scheme.

There are many model-based approaches to CF. Some are based on
linear algebra (SVD, PCA, or
Eigenvectors)~\cite{FactorAnalysis,Competitive,EigentasteFull,Fuzzy,DimensionalityReduction,incrementalsvd};
 or on techniques
borrowed more directly from Artificial Intelligence such as Bayes
methods, Latent Classes, and Neural
Networks~\cite{Neural,Empirical,hofmann:latent}; or on
clustering~\cite{Nannen:Thesis:2003,chee01rectree}. In comparison
to memory-based schemes, model-based CF algorithms are typically
faster at query time though they might have expensive learning or
updating phases. Model-based schemes can be preferable to
memory-based schemes when query speed is crucial.

We can compare our predictors with certain types of predictors
described in the literature in the following algebraic terms. Our
predictors are of the form $f(x) = x+b$, hence the name ``slope
one'', where $b$ is a constant and $x$ is a variable representing
rating values. For any pair of items, we attempt to find the best
function $f$ that predicts one item's ratings from the other
item's ratings. This function could be different for each pair of
items. A CF scheme will weight the many predictions generated by
the predictors. In~\cite{Item-based}, the authors considered the
correlation across pairs of items and then derived weighted
averages of the user's ratings as predictors. In the simple
version of their algorithm, their predictors were of the form
$f(x)=x$. In the regression-based version of their algorithm,
their predictors were of the form $f(x)= ax+b$.
In~\cite{Regression-based}, the authors also employ predictors of
the form $f(x)=ax+b$. A natural extension of the work in these two
papers would be to consider predictors of the form $f(x)=
ax^2+bx+c$. Instead, in this paper, we use naïve predictors of the
form $f(x)= x+b$. We also use naïve weighting. It was observed
in~\cite{Item-based} that even their regression-based $f(x)= ax+b$
algorithm didn't lead to large improvements over memory-based
algorithms. It is therefore a significant result to demonstrate
that a predictor of the form $f(x)=x+b$ can be competitive with
memory-based schemes.

\section{CF Algorithms}
\label{Schemes} We propose three new CF schemes, and contrast our
proposed schemes with four reference schemes: \textsc{Per User
Average}, \textsc{Bias From Mean}, \textsc{Adjusted Cosine
Item-Based}, which is a model-based scheme, and the
\textsc{Pearson} scheme, which is representative of memory-based
schemes.

\subsection{Notation}
We use the following notation in describing schemes. The ratings
from a given user, called an \textit{evaluation}, is
represented as an incomplete array $u$, where $u_i$ is the rating of this
user gives to item $i$. The subset of the set of items consisting of
all those items which are rated in $u$ is
$S(u)$. The set of all evaluations in the training set is $\chi$. The number of
elements in a set $S$ is $card(S)$. The average of ratings in an
evaluation $u$ is denoted $\bar{u}$. The set $S_i(\chi)$ is the set of all
evaluations $u\in \chi$ such that they contain item $i$ ($i\in
S(u)$). Given two evaluations $u,v$, we define the scalar product
$\langle u,v \rangle$ as $\sum_{i\in S(u)\cap S(v)} u_i v_i $.
Predictions, which we write $P(u)$, represent a vector where each component is
the prediction corresponding to one item: predictions depend implicitly
on the training set $\chi$.

\subsection{Baseline Schemes}
\label{ssec:base}One of the most basic prediction algorithms is
the \textsc{Per~User~Average} scheme given by the equation $P(u) =
\bar{u}$. That is, we predict that a user will rate everything
according to that user's average rating. Another simple scheme is
known as \textsc{Bias From Mean} (or sometimes
\textsc{Non~Personalized}~\cite{NonPersonalized}). It is given by
\[P(u)_i=\bar{u} + \frac{1}{card(S_i(\chi))} \sum_{v\in S_i(\chi)} v_i-\bar{v} .\]
That is, the prediction is based on the user's average plus the
average deviation from the user mean for the item in question
across all users in the training set. We also compare to the
item-based approach that is reported to work
best~\cite{Item-based}, which uses the following adjusted cosine
similarity measure, given two items $i$ and $j$:
\[\textrm{sim}_{i,j}= \frac{\sum_{u\in S_{i,j}(\chi)} (u_i-\bar{u})(u_j-\bar{u})}
{\sqrt{\sum_{u\in S_{i,j}(\chi)} (u_i-\bar{u})^2 \sum_{u\in S_{i,j}(\chi)} (u_j-\bar{u})^2}}\]
The prediction is obtained as a weighted sum of these measures thus:
\[P(u)_i=  \frac{\sum_{j\in S(u)} \vert \textrm{sim}_{i,j} \vert (\alpha_{i,j} u_j + \beta_{i,j})}{\sum_{j\in S(u)} \vert \textrm{sim}_{i,j} \vert}\]
where the regression coefficients $\alpha_{i,j},\beta_{i,j}$ are chosen so as to
minimize $\sum_{u\in S_{i,j}(u)}(\alpha_{i,j} u_j \beta_{i,j} - u_i)^2$ with $i$ and $j$ fixed.

\subsection{The \large{\textbf{\textsc{Pearson}}} \textbf{Reference Scheme}}
Since we wish to demonstrate that our
schemes are comparable in predictive power to memory-based schemes, we choose to implement one
such scheme as representative of the class, acknowledging that there are many documented schemes of this type.
Among the most popular and accurate memory-based schemes is the \textsc{Pearson}
scheme~\cite{GroupLens1994}. It takes the form of a weighted sum over all users in $\chi$
\[P(u)_i=\bar{u} + \frac{\sum_{v\in S_i(\chi)} \gamma(u,v) (v_i-\bar{v})}{\sum_{v\in S_i(\chi)} \vert \gamma(u,v) \vert }  \]
where $\gamma$ is a similarity measure computed from Pearson's correlation:
\[Corr(u,w)=\frac{\langle u-\overline{u},w-\bar{w} \rangle }
{\sqrt{\sum_{i\in S(u) \cap S(w)} (u_i - \overline{u})^2
\sum_{i\in S(u) \cap S(w)} (w_i - \overline{w})^2 }}.\]
Following~\cite{Empirical,NonPersonalized}, we set
\[
\gamma(u,w)=Corr(u,w)\left \vert Corr(u,w)\right \vert^{\rho-1}\]
 with $\rho=2.5$, where $\rho$ is the Case Amplification power.
 Case Amplification reduces noise in the data:
 if the correlation is high, say $Corr=0.9$, then it remains high
 ($0.9^{2.5}\cong 0.8$) after Case Amplification whereas if it is low, say $Corr=0.1$, then it
 becomes negligible ($0.1^{2.5}\cong 0.003$).
Pearson's correlation together with Case Amplification is shown to
be a reasonably accurate memory-based scheme for CF
in~\cite{Empirical} though more accurate schemes exist.

\subsection{The \large{\textbf{\textsc{Slope~One}}} \textbf{Scheme}}
\label{ssec:slopeone} The slope one schemes take into account both
information from other users who rated the same item (like the
\textsc{Adjusted Cosine Item-Based}) and from the other items
rated by the same user (like the \textsc{Per User Average}).
However, the schemes also rely on data points that fall neither in
the user array nor in the item array (e.g. user $A$'s rating of
item $I$ in Fig.~\ref{figsimple}), but are nevertheless important
information for rating prediction. Much of the strength of the
approach comes from data that is \emph{not} factored in.
Specifically, only those ratings by users who have rated some
common item with the predictee user and only those ratings of
items that the predictee user has also rated enter into the
prediction of ratings under slope one schemes.

Formally, given two evaluation arrays $v_i$ and $w_i$ with $i =
1,\ldots, n$, we search for the best predictor of the form $f(x) =
x + b $ to predict $w$ from $v$ by minimizing $\sum_i ( v_i +
b - w_i)^2$. Deriving with respect to $b$ and setting the
derivative to zero, we get $b = \frac{\sum_i w_i -  v_i }{n}$. In
other words, the constant $b$ must be chosen to be the average
difference between the two arrays. This result motivates the
following scheme.

Given a training set $\chi$, and any two items $j$ and $i$ with ratings
$u_j$ and $u_i$ respectively in some user evaluation $u$ (annotated as $u$$\in$$S_{j,i}(\chi)$), we consider the
average deviation of item $i$ with respect to item $j$ as:
\[\textrm{dev}_{j,i}=\sum_{u\in S_{j,i}(\chi)}\frac{u_j-u_i}{card(S_{j,i}(\chi))}.\]
Note that any user evaluation $u$ not containing both $u_j$ and $u_i$ is not
included in the summation.
The symmetric matrix defined by $\textrm{dev}_{j,i}$ can be computed once and updated
quickly when new data is entered.

Given that $\textrm{dev}_{j,i} +u_i$ is a prediction for $u_j$ given $u_i$,
a reasonable predictor might be the average of all such predictions
\[P(u)_j =
\frac{1}{card(R_j)} \sum_{i\in R_j} (\textrm{dev}_{j,i} +u_i) \] where $R_j
= \{ i | i\in S(u), i\neq j, card(S_{j,i}(\chi)) > 0 \}$ is the set of all relevant
items. There is an approximation that can simplify the calculation of this prediction.
For a dense enough data set where almost all pairs of items have
ratings, that is, where $card(S_{j,i}(\chi)) > 0$ for almost all
$i,j$, most of the time $R_j = S(u)$ for  $j\notin S(u)$ and
$R_j = S(u)-\{j\}$ when $j\in S(u)$. Since $\bar{u} = \sum_{i\in
S(u)} \frac{u_i}{card(S(u))} \simeq \sum_{i\in R_j}
\frac{u_i}{card(R_j)}$ for most $j$, we can simplify the
prediction formula for the \textsc{Slope~One} scheme to
\[P^{S1}(u)_j = \bar{u} + \frac{1}{card(R_j)} \sum_{i\in R_j} \textrm{dev}_{j,i}. \]
It is interesting to note that our implementation of
\textsc{Slope~One} doesn't depend on how the user rated individual
items, but only on the user's average rating and crucially on
which items the user has rated.

\subsection{The \large{\textbf{\textsc{Weighted~Slope~One}}} \textbf{Scheme}}
\label{ssec:weightSone}One of the drawbacks of \textsc{Slope~One} is that the number of
ratings observed is not taken into consideration. Intuitively, to predict user
$A$'s rating of item $L$ given user $A$'s rating of items $J$ and
$K$, if 2000 users rated the pair of items $J$ and $L$ whereas only 20 users
rated the pair of items $K$ and $L$, then user $A$'s rating of item $J$ is
likely to be a far better predictor for item $L$ than user $A$'s
rating of item $K$ is. Thus, we define the
\textsc{Weighted~Slope~One} prediction as the following weighted
average
\[P^{wS1}(u)_j =
\frac{ \sum_{i\in S(u)-\{j\}} (\textrm{dev}_{j,i} +u_i) c_{j,i}
}{\sum_{i\in S(u)-\{j\}} c_{j,i}}
 \]
 where $c_{j,i}=card(S_{j,i}(\chi))$.

\subsection{The \large{\textbf{\textsc{Bi-Polar~Slope~One}}} \textbf{Scheme}}
\label{ssec:bipolar}

While weighting served to favor frequently occurring rating
patterns over infrequent rating patterns, we will now consider
favoring another kind of especially relevant rating pattern. We
accomplish this by splitting the prediction into two parts. Using
the \textsc{Weighted Slope One} algorithm, we derive one
prediction from items users liked and another prediction using
items that users disliked.

Given a rating scale, say from 0 to 10, it might seem reasonable
to use the middle of the scale, 5, as the threshold and to say
that items rated above 5 are liked and those rated below 5 are
not. This would work well if a user's ratings are distributed
evenly. However, more than 70\% of all ratings in the EachMovie
data are above the middle of the scale. Because we want to support
all types of users including balanced, optimistic, pessimistic,
and bimodal users, we apply the user's average
% AEM: Daniel, did we ever try MEDIAN instead of AVERAGE?
as a threshold between the users liked and disliked items. For example,
optimistic users, who like every item they rate, are assumed to
dislike the items rated below their average rating. This threshold
ensures that our algorithm has a reasonable number of liked and
disliked items for each user.

Referring again to Fig.~\ref{figsimple}, as usual we base our
prediction for item $J$ by user $B$ on deviation from item $I$ of
users (like user $A$) who rated both items $I$ and $J$. The
\textsc{Bi-Polar Slope One} scheme restricts further than this the
set of ratings that are predictive. First in terms of items, only
deviations between two liked items or deviations between two
disliked items are taken into account. Second in terms of users,
only deviations from pairs of users who rated both item $I$ and
$J$ and who share a like or dislike of item $I$ are used to
predict ratings for item $J$.

The splitting of each user into user likes and user dislikes
effectively doubles the number of users. Observe, however, that
the bi-polar restrictions just outlined necessarily reduce the
overall number of ratings in the calculation of the predictions.
Although any improvement in accuracy in light of such reduction
may seem counter-intuitive where data sparseness is a problem,
failing to filter out ratings that are irrelevant may prove even
more problematic. Crucially, the \textsc{Bi-Polar Slope One}
scheme predicts nothing from the fact that user $A$ likes item $K$
and user $B$ dislikes this same item $K$.

Formally, we split each evaluation in $u$ into two sets of rated
items: $S^{like}(u)= \{i\in S(u)| u_i > \bar u \}$ and
$S^{dislike}(u)= \{i\in S(u)| u_i < \bar u \}$. And for each pair
of items $i,j$, split the set of all evaluations $\chi$ into
$S_{i,j}^{like}=\{u\in \chi | i,j \in S^{like}(u)\}$ and
$S_{i,j}^{dislike}=\{u\in \chi | i,j \in S^{dislike}(u)\}$. Using
these two sets, we compute the following deviation matrix for
liked items as well as the derivation matrix
${dev}_{j,i}^{dislike}$.
\[\textrm{dev}_{j,i}^{like}=\sum_{u\in S_{j,i}^{like}(\chi)}\frac{u_j-u_i}{card(S_{j,i}^{like}(\chi))},\]
The prediction for rating of item $j$ based on rating of item $i$
is either $p_{j,i}^{like}=\textrm{dev}_{j,i}^{like} + u_i$ or
$p_{j,i}^{dislike}=\textrm{dev}_{j,i}^{dislike}+ u_i$ depending on
whether $i$ belongs to $S^{like}(u)$ or $S^{dislike}(u)$
respectively. The \textsc{Bi-Polar Slope One} scheme is given by
\[P^{bpS1}(u)_j = \frac{
\begin{array}{l}
\sum_{ i\in S^{like}(u)-\{j\}} p_{j,i}^{like}  c_{j,i}^{like}
 \\
+\sum_{i\in S^{dislike}(u)-\{j\}} p_{j,i}^{dislike} c_{j,i}^{dislike}
\end{array}
}{
    \sum_{i\in S^{like}(u)-\{j\}} c_{j,i}^{like}+
    \sum_{i\in S^{dislike}(u)-\{j\}} c_{j,i}^{dislike}
}
 \]
where the weights
 $c_{j,i}^{like}= card(S_{j,i}^{like})$
 and $c_{j,i}^{dislike}= card(S_{j,i}^{dislike})$ are similar to the
 ones in the \textsc{Weighted Slope One} scheme.

\section{Experimental Results}
\label{ssec:Results} The effectiveness of a given CF algorithm can
be measured precisely. To do so, we have used the All But One Mean
Average Error (MAE)~\cite{Empirical}. In computing MAE, we successively hide ratings
one at a time from all evaluations in the test set while predicting the hidden
rating, computing the average error we make in the prediction.
Given a predictor $P$ and an
evaluation $u$ from a user, the error rate of $P$ over a set
of evaluations $\chi'$, is given by
\[MAE = \frac{1}{card(\chi')}\sum_{u\in \chi'}\frac{1}{card(S(u))}\sum_{i\in S(u)}\vert P(u^{(i)})-u_i \vert \]
where $u^{(i)}$ is user evaluation $u$ with that user's rating of
the $i$th item, $u_i$, hidden.

We test our schemes over the EachMovie data set made available by
Compaq Research and over the Movielens data set from the Grouplens
Research Group at the University of Minnesota. The data is collected
from movie rating web
sites where ratings range from $0.0$ to $1.0$ in increments of
$0.2$ for EachMovie and from 1 to 5 in increments of 1 for Movielens.
Following~\cite{NonPersonalized,LemireIR2003}, we used
enough evaluations to have a total of 50,000 ratings as a training
set ($\chi$) and an additional set of evaluations with a total of
at least 100,000 ratings as the test set ($\chi'$).
%, and we repeated the process
%6 times over different pairs $\chi,\chi'$ for each data set keeping
%only the average.
When predictions fall outside the range of allowed ratings for the
given data set, they are corrected accordingly: a prediction of
1.2 on a scale from 0 to 1 for EachMovie is interpreted as a prediction of 1.
Since Movielens has a rating scale 4 times larger than EachMovie,
MAEs from Movielens were divided by 4 to make the results directly comparable.

\begin{table}\begin{footnotesize}
\begin{center}\begin{tabular}{|c|c|c|}
\hline Scheme
%&  MAE (>20 ratings)
& EachMovie & Movielens
\\
\hline \hline
\textsc{Bi-Polar~Slope~One}  & 0.194 & 0.188 \\
\hline
\textsc{Weighted~Slope~One}  & 0.198 & 0.188\\
\hline
\textsc{Slope~One}  & 0.200 & 0.188 \\
\hline \hline
\textsc{Bias From Mean}  & 0.203 & 0.191 \\
\hline
\textsc{Adjusted Cosine Item-Based}  & 0.209 & 0.198 \\
\hline
\textsc{Per User Average}  & 0.231 & 0.208 \\
\hline \hline
\textsc{Pearson}  & 0.194 & 0.190\\
\hline
\end{tabular}\end{center}
\end{footnotesize}

\caption{\label{tableresults}All Schemes Compared: All But One
Mean Average Error Rates for the EachMovie and Movielens data sets, lower
is better.}
\end{table}

The results for the various schemes using the same error measure
and over the same data set are summarized in
Table~\ref{tableresults}. Various subresults are highlighted in
the Figures that follow.

Consider the results of testing various baseline schemes. As
expected, we found that \textsc{Bias~From~Mean} performed the best
of the three reference baseline schemes described in
section~\ref{ssec:base}. Interestingly, however, the basic
\textsc{Slope One} scheme described in section~\ref{ssec:slopeone}
had a higher accuracy than \textsc{Bias~From~Mean}.

The augmentations to the basic \textsc{Slope One} described in
sections~\ref{ssec:weightSone} and~\ref{ssec:bipolar} do improve
accuracy over EachMovie. There is a small difference between
\textsc{Slope~One} and \textsc{Weighted Slope One} (about 1\%).
Splitting dislike and like ratings improves the results 1.5--2\%.

Finally, compare the memory-based \textsc{Pearson} scheme on the
one hand and the three slope one schemes on the other. The slope
one schemes achieve an accuracy comparable to that of the
\textsc{Pearson} scheme. This result is sufficient to support our
claim that slope one schemes are reasonably accurate despite their
simplicity and their other desirable characteristics.

\section{Conclusion}
This paper shows that an easy to implement CF model based on
average rating differential can compete against more expensive
memory-based schemes. In contrast to currently used schemes, we
are able to meet 5 adversarial goals with our approach. Slope One
schemes are easy to implement, dynamically updateable, efficient
at query time, and expect little from first visitors while having
a comparable accuracy (e.g. 1.90 vs. 1.88 MAE for MovieLens) to
other commonly reported schemes. This is remarkable given the
relative complexity of the memory-based scheme under comparison. A
further innovation of our approach are that splitting ratings into
dislike and like subsets can be an effective technique for
improving accuracy. It is hoped that the generic slope one
predictors presented here will prove useful to the CF community as
a reference scheme.

Note that as of November 2004, the \textsc{Weighted~Slope~One} is the 
collaborative filtering algorithm used by the Bell/MSN Web site 
inDiscover.net.

\bibliographystyle{plain}
\bibliography{linear}
\end{document}